\documentclass{article}
\usepackage[T1]{fontenc}
\usepackage[utf8]{inputenc}
\usepackage{amsmath,amssymb}
\usepackage[english]{babel}
\usepackage{url}
\usepackage{hyperref}
\begin{document}

%\begin{center}
%\tiny{Ethereum Smart Contract Security:\\ 
%\url{https://etherscan.io/address/0xe342f428cc07e9228b64095ec0b0cd329dea1039}\\
%February 7, 2019 - v2.0\\
%Upload data: March 18, 2019
%}
%\end{center}

\begin{center}
\textbf{\Large{Delta - new logic programming language.\\
	$\&$\\ 
	Delta-methodology for p-computable programs on Turing Complete Languages.\\}}
\end{center}

\begin{center}
\large{Andrey Nechesov}
\end{center}

\begin{center}
Russia, Novosibirsk, Academgorodok\\ 
Sobolev institute of mathematics.\\
Email: nechesoff@gmail.com\\
Telegram: @nechesoff\\
Skype: Nechesov
\end{center}

\textbf{Annotation:} In paper describes the new logic programming language \textit{Delta}, which have a many good properties. Delta-programs is p-computable, verifiable and can translation on other languages. Also we describe the Delta-methodology for constructing p-computable programs in high-level languages such as PHP, Java, JavaScript, C++, Pascal, Delphi, Python, Solidity and other. We would like to especially note the use of the Delta methodology for creating \textit{Smart Contracts} and for \textit{Internet of things}.\\

\textbf{Keywords:} Delta, Delta language, dynamic logic, dynamic model, semantic programming, logic language, Delta-methodology, polynomial computability, turing complete languages, polynomial time, p-computable program, blockchain, smart contracts, cryptocurrency, Ethereum, Bitcoin, internet of things, IoT.\\

\textbf{Introduction}\\

In this paper introduced new logic programming language Delta for building logical Delta-programs based on the theory of semantic programming developed by Ershov, Goncharov and Sviridenko\cite{b_sigma_prog}\cite{b_ges} in the 70s-80s of the last century.\\
In this paper we change the concept of the formula and define \textit{D-formulas(or Delta programs)} are special list-formulas. Then we define the execution of a program how is the process of checking truth D-formula on a dynamic model.\\

Polynomiality is the main advantage of the Delta programs, which allows not only to build Delta programs, but also use Delta methodology for creation programs in high-level languages.\\

Main idea our paper consider program how list-formula from another formulas on dynamic models. And we created by iterations new Delta-programs use simple base formulas for this. Also we entered a dynamic models how models where we save final values of variables when check formula on this model.\\

\textbf{1. Semantic programming}\\

The main idea of semantic programming is to consider the program as a formula on a suitable model and reduce the execution of the program to the truth checking formula on the model. Ershov, Goncharov, Sviridenko\cite{b_sigma_prog} in their works suggested using the hereditarily-finite super structure $HW(\mathfrak{M})$. They added in base set of model new elements - lists and add new $Lisp$ functions and relations.\\

We will use some of the list-functions in our article:\\

1) $nil$ - the empty list constant\\

2) $head$ - last element in non-empty list or $nil$ otherwise.\\

3) $tail$ - the list retrieved from the base non-empty list by deleting the last element or $nil$ otherwise.\\

4) $cons(l_1,l_2)$ - add list $l_2$ how new last element in $l_1$\\

5) $conc(l_1,l_2)$ - concatenation of 2 lists: $<l_1,l_2>$\\

6) $l\in w$ - where $l$ - element of list $w$\\

7) $l\subseteq w$ - where $l$ to be the beginning of the list $w$\\

and we add some next operations:\\

8) $addValue(l,<x,a>)$ - delete any element with view $<x,b>$ for some $b$ from list $l$ and add $<x,a>$ in list $l$  how last elements.\\

9) $addValues(l,<x_1,a_1>,...,<x_n,a_n>)$ - delete any elements with view $<x_i,b_i>$ for some $b_i$ from list $l$ and add $<x_i,a_i>$ in list $l$ how last elements.\\

In semantic programming we can define new types of objects with $\Delta_0^p-$enrichments and add this objects in base set.\cite{b_nechesov}. Set of this objects will be a p-computable. This helps us to extend the set of types of variables in the model, while not going beyond the polynomiality.\\

In \cite{b_logic2018} we defined $\Delta_0$-boundary terms, which can extend our formulas and new formulas extension will be conservative. The boundary property will be used very often in our article, because the polynomiality is very closely with it. All our new constructions will be boundary.\\

All this methods and extensions we can use in our Delta-methodology, which we describe below.\\

\textbf{2. Delta abstract logic language and Dynamics models.}\\

When we create a program, it is very important that in the process of executing the program code, we save our calculations. In the program code, assigning a variable a value, we store this information. Therefore, when creating a logical program, it is very important to store the values of variables. Standart logic models not enough good do this. In this chapter we entered a new abstract logic lanquage \textit{Delta} and new type of models - Dynamics models. On this models we extended formulas with new operators and symbols, and now we can easy save calculated variable values. Dynamic model created an initialized list of variables and add another values of variables for new D-predicates and 
D-functions. Any logical program we consider how list of $<\Phi_1,...,\Phi_n>$ of D-formulas $\Phi_i$ on our dynamic model $D(\mathfrak{M})_E$. When we check formula $\Phi_i$ on truth $D(\mathfrak{M})\models\Phi_i$, our model $D(\mathfrak{M})_E$ can change own internal parameters with formula operator $\Gamma_{\Phi_i}(E)$ and can change signature  $\sigma$: add new predicate or function symbols.\\

Let $\mathfrak{M}-$ polinomial model signature $\sigma$.\\

\textbf{Denotement:} Dynamic model: 
\begin{center}
$D(\mathfrak{M})_E = <\mathfrak{M},E>$, where $E$ - trace list of sets of initialisation variables pairs $<variable,value>$ (on start $E=nil$)\\
\end{center}
with signature $\sigma^*=\sigma$ on start, but can enrich another predicates and functions symbols.\\ 

\textbf{Denotement:} Formula $\Phi(\overline{x},\overline{y})\ -$ $\Delta_0^p$-formula, if truth checking formula $\Phi$ is p-computable algorithm, from incoming variables $x_i = a_i$ and this algorithm also find values for outcoming variables $\overline{y}$.\\

\textbf{Denotement} \textit{Boundary $\Delta_0^p$-formula} $\Phi(\overline{x},\overline{y})$ it's $\Delta_0^p$-formula, , where 
\begin{center}
$\exists C\ \exists p\ \forall x_i\ |y_i|\leq C*(|x_1|+...|x_n|)^p$.\\ 
\end{center}

\textbf{Denotement:} Formula $\Phi(\overline{x},\overline{y})\ -$ C-p-$\Delta_0^p$-formula, if truth checking formula $\Phi$ is p-computable algorithm with contant $C$ and power of $p$.\\

\textbf{Denotement:} $\mathfrak{M}\ -$ $\Delta_0^p$-model, if $\mathfrak{M}$ - p-computable model.\\

Let $F = \{\Phi_i\}$, $i\in N$ - countably or finite family boundary of C-p-$\Delta_0^p$-formulas on C-p-$\Delta_0^p$-model $\mathfrak{M}$.\\

\textbf{Denotement:} Family $F$, defined above, is boundary C-p-$\Delta_0^p$-family.\\

%\textbf{Definition:} We say that C-p-$\Delta_0^p$-family $F$ is \textit{generating family for} $\Delta_0^p-$predicate $P(\overline{x},\overline{y})$, if exists $\Delta_0^p$-function $\alpha(\overline{x})$ which for each $\overline{a}$ create formula $\Phi_{i(\overline{a})}(\overline{a},\overline{y})\in F$ and for any $\Phi(\overline{x},\overline{y})\in F$ exist $\overline{b}$, $\alpha(\overline{b}) = \Phi(\overline{b},\overline{y})$. This $\Delta_0^p$-predicates we call D-predicates.\\

Let $\mathfrak{M}$ - $\Delta_0^p$-model signature $\sigma$.\\

Let $D(\mathfrak{M})_E$-dynamic model signature $\sigma^*$, where on start $E=nil$\\

\textit{Inductively define D-Terms on dynamic model $D(\mathfrak{M})_E$ with signature $\sigma^*$:}\\

1) if $c$ - constant, then $c$ - D-term\\

2) if $x$ - variable, then $x$ - D-term\\

3) if $f\in\sigma^*$ - n-th place functional symbol, then $f(t_1(\overline{x}),...,t_n(\overline{x}))$ - D-term, where $t_i$ - D-terms\\

\textit{Inductively define \textit{D-formulas} on dynamic model $D(\mathfrak{M})_E$:}\\

1) Any quantifer free formula of signature $\sigma^*$ is D-formula\\

2) "$y:=t(\overline{x})$"\ - it's D-formula(\textit{"assignment"\ operator}), where $t-$D-term .\\

3) $COPY$ operator: analoge FOR in programming languages but we requare what all outcoming variables $\overline{y}$ was a boundary:\\
"$COPY(\Phi(\overline{x},\overline{y}),n) = <\Phi,...,\Phi>$"\, copy the formula n-times $\Phi$.\\

 for $\Psi(\overline{x},\overline{y}): <\Phi,...,\Phi>$ $\exists C,p \forall n\in N |y_i|\leq C*(|x_1|+...|x_n|+n)^p$\\

4) $If$ operator:
\begin{equation*}If(\Psi(\overline{x},\overline{y}),\Phi_1(\overline{x},\overline{y}),\Phi_2(\overline{x},\overline{y})) = 
	 \begin{cases}
	   \Phi_1,\ D(\mathfrak{M})_E\models \Psi(\overline{x},\overline{y}) \\			   
	   \Phi_2,\ else 
	 \end{cases}
	\end{equation*}

5) $\Delta_0^p-$operator $\alpha$: $\alpha(\overline{x}) = \Phi(\overline{x},\overline{y})$, $\Phi\in F$, 
$F-$boundary $\Delta_0^p-$family and $\alpha\in\sigma^*-$special functional symbol.
\begin{center}
$\alpha(x)-$D-formula. 
\end{center}

6) "predicate $P(\overline{x},\overline{y}):\Phi(\overline{x},\overline{y})$"\ is D-formula (or \textit{D-predicate}), where $P\notin\sigma^*$ and $\Phi(\overline{x},\overline{y})$ -D-formula.\\

7) "function $f(\overline{x})\ return\ y:\Phi(\overline{x}, y)$"\ is D-formula ($f(\overline{x})$ is \textit{D-function}), if $(f(\overline{x}),y)$ - D-predicate and $f\notin\sigma^*$,.\\ 

8) "$return$"\ - D-formula\\

9) List of formulas "$<\Phi_1,...,\Phi_n>$"\ is D-formula, all $\Phi_i$ - D-formulas.\\

% - stop program executing, or return predicate or function outcoming values;\\

\textit{Inductively define truth checking D-formula on $D(\mathfrak{M})_E$ and how model $D(\mathfrak{M})_E$} is changed\\

1) If $\Phi(\overline{x})$ - quantifier free formula signature $\sigma^*$, then
\begin{center}
$D(\mathfrak{M})_E\models \Phi(\overline{x})\Leftrightarrow \mathfrak{M}\models \Phi(\overline{a})$
\end{center}
where all $<x_i,a_i>\in head(E)$.\\

2) Operator $COPY$:

\begin{center}
$D(\mathfrak{M})_E\models COPY(\Phi(\overline{x},\overline{y}),z) \Leftrightarrow D(\mathfrak{M})_E\models <\Phi(\overline{x},\overline{y}),...,\Phi(\overline{x},\overline{y})>$
\end{center}
where all $<x_i,a_i>\in head(E)$ and $<z,n>\in head(E)$, $n\in N$ and for $\Psi(\overline{x},\overline{y}): <\Phi(\overline{x},\overline{y}),...,\Phi(\overline{x},\overline{y})>$ we requare for all $y_i$ in $\Psi$ is boundary from $\overline{x}$ and $n$.\\

3) Operator $IF$:

\begin{equation*}D(\mathfrak{M})_E\models If(\Psi,\Phi_1,\Phi_2) \Leftrightarrow  
	 \begin{cases}
	   D(\mathfrak{M})_{\Gamma_\Psi(E)}\models\Phi_1,\ if\ D(\mathfrak{M})_E\models\Psi(\overline{x},\overline{y})\\			   
	   D(\mathfrak{M})_{E}\models\Phi_2,\ else 
	 \end{cases}
	\end{equation*}

{\footnotesize $\Gamma_{\Psi}(E) = cons(tail(E),addValues(head(E),<y_1,b_1>,...,<y_k,b_k>))$\\
and all $<x_i,a_i>\in head(E)$, $i\in [1,...,n]$\\ 
}

4) Operator $\alpha$:

\begin{center}
$D(\mathfrak{M})_E\models\alpha(\overline{x})\Leftrightarrow D(\mathfrak{M})_E\models \Phi(\overline{x},\overline{y})$
\end{center}

{\footnotesize where all $<x_i,a_i>\in E$ for some $a_i$ and $\Phi\in F$ where  $F$ boundary-C-p-$\Delta_0^p-$family}\\

5) Let $f\in\sigma -$n-place functional symbol
\begin{center}
$D(\mathfrak{M})_E\models y:=f(\overline{x})\Leftrightarrow D(\mathfrak{M})_{\Gamma_{(f(\overline{x}),y)}(E)}\models true$\\
\end{center}
{\footnotesize $\Gamma_{(f(\overline{x}),y)}(E) = cons(tail(E),addValue(head(E),<y,f(\overline{a})>))$\\
and all $<x_i,a_i>\in head(E)$, $i\in [1,...,n]$\\ 
}

6) Let $f\in\sigma -$n-place D-functional symbol, where $(f(\overline{x}),y)$ - D-predicate
\begin{center}
$D(\mathfrak{M})_E\models z:=f(\overline{x})\Leftrightarrow D(\mathfrak{M})_{\Gamma_{(f(\overline{x}),z)}(E)}\models \alpha(\overline{a})$
\end{center}
{\footnotesize $\Gamma_{(f(\overline{x}),z)}(E) = cons(cons(E,<<z,y>>),<<x_1,a_1>,...,<x_n,a_n>>)$}\\ 
{\footnotesize and all $<x_i,a_i>\in head(E)$, $i\in [1,...,n]$}\\

7) Let $t_1,...,t_n$ - D-terms of signature $\sigma^*$, and $f\in\sigma^*$ - n-place functional symbol.
\begin{center}
$D(\mathfrak{M})_E\models y:=f(t_1(\overline{x}),...,t_n(\overline{x}))$ \\
$\Leftrightarrow$\\
$D(\mathfrak{M})_E\models <z_1:=t_1(\overline{x}),...,z_n:=t_n(\overline{x}),y:=f(z_1,...,z_n)>$\\ 
{\footnotesize and all $<x_i,a_i>\in head(E)$ for all $i\in [1,...,n]$}\\
$\Leftrightarrow$\\
$D(\mathfrak{M})_{\Gamma_{(z_n,t_n)}(...\Gamma_{(z_1,t_1)}(E)...)}\models y:=f(z_1,...,z_n)$
\end{center}
{\footnotesize $\Gamma_{(z_n,t_n)}(...\Gamma_{(z_1,t_1)}(E)...) = cons(tail(E),addValues(head(E), <z_1,t_1>,...,<z_n,t_n>))$}\\

8) Let $P(\overline{x},\overline{y})$ - boundary $\Delta_0^p$-predicate of signature $\sigma$.
\begin{center}
	$D(\mathfrak{M})_E\models P(\overline{x},\overline{y}) \Leftrightarrow D(\mathfrak{M})_{\Gamma_{P}(E)} \models true$
	\end{center}
	{\footnotesize $\Gamma_P(E)=cons(tail(E),addValues(head(E),<y_1 = b_1>,...,<y_k=b_k>))\ \&$ all $<x_i,a_i>\in head(E)$}\\

9) Let $P(\overline{x},\overline{y})$ - D-predicate of signature $\sigma^*$.
\begin{center}
	$D(\mathfrak{M})_E\models P(\overline{x},\overline{z}) \Leftrightarrow D(\mathfrak{M})_{\Gamma_P(E)} \models \Phi(\overline{x},\overline{y})$
\end{center}
{\footnotesize $\Gamma_P(E) = cons(cons(tail(E),<<z_1,y_1>,...,<z_n,y_n>>),<<x_1,a_1>,...,<x_n,a_n>>)$
and all $<x_i,a_i>\in head(E)$, $i\in[1,...,n]$}\\

10) $D(\mathfrak{M})_E\models <\Phi_1,...,\Phi_n> \Leftrightarrow D(\mathfrak{M})_E\models\Phi_1$ and $D(\mathfrak{M})_{\Gamma_{\Phi_1}(E)}\models <\Phi_2,...,\Phi_n>$\\

11) $D(\mathfrak{M})_E\models return \Leftrightarrow D(\mathfrak{M})_{\Gamma_{return}(E)}\models true$\\

{\footnotesize $\Gamma_{return}(E)=cons(tail(E), addValue(head(E),<return,1>))$}\\

12) $D(\mathfrak{M})_{\Gamma_{return}(E)}\models <\Phi_1,...,\Phi_n> \Leftrightarrow D(\mathfrak{M})_{\Gamma_{return}(E)}\models <\Phi_2,...,\Phi_n>$\\

13) $D(\mathfrak{M})_{\Gamma_{return}(E)}\models nil \Leftrightarrow D(\mathfrak{M})_{\Gamma_{nil}(\Gamma_{return}(E))}\models true$\\

{\footnotesize $\Gamma_{nil}(\Gamma_{return}(E))=cons(tail(tail(tail(\Gamma_{return}(E)))), addValues(head(tail(tail(\Gamma_{return}(E)))),<z_1=b_1>,...,<z_k,b_k>))$}\\

14) $D(\mathfrak{M})_{E}\models <true,\Phi_1,...,\Phi_n> \Leftrightarrow D(\mathfrak{M})_{E}\models <\Phi_1,...,\Phi_n>$\\

15) $D(\mathfrak{M})_{E}\not\models <\Phi_1,...,\Phi_n> \Leftrightarrow \exists i\ D(\mathfrak{M})_{\Gamma_{\Phi_{i-1}}(...(\Gamma_{\Phi_1}(E)...)}\not\models \Phi_i$\\

16) $D(\mathfrak{M})_{E}\models predicate\ P(\overline{x},\overline{y}):\Phi(\overline{x},\overline{y}) \Leftrightarrow P\notin\sigma^*$ and $\Phi$-D-formula\\

17) $D(\mathfrak{M})_{E}\models function\ f(\overline{x})\ return\ y:\Phi(\overline{x},y) \Leftrightarrow f\notin\sigma^*$ and $\Phi$-D-formula\\

\textbf{3. Delta programs}\\

When we define the concept of a D-formulas and dynamic models, we can introduce the concept of a delta program and describe the process of computation. Calculation of outgoing values in a delta-program and checking truth of D-formula on dynamic model is equivalent concepts.\\

\textbf{Definition: }\textit{\textit{Delta-program} it's D-formula on dynamic model $D(\mathfrak{M})_E$}\\

\textbf{Definition: }\textit{Delta-program} $\Phi$ is $\Delta_0^p$-program, if $\Phi -\Delta_0^p-$D-formula\\

\textbf{Definition: }\textit{\textit{Process of computation of Delta-program} it's process of truth checking this how D-formula on dynamic model $D(\mathfrak{M})_E$}\\

\textbf{Definition: } \textit{Atomic D-formula} $\Phi$ it's formula have a next one view:\\ 
1) $y:=f(\overline{x})$, where $f\in \sigma$\\
2) $P(\overline{x},\overline{y})$, where $P\in \sigma$\\
3) $return$\\

Inductively define rank $r(\Phi)$ for any D-formula on dynamic-model $D(\mathfrak{M})_E$:\\

1) $r(\Phi) = 0$, if $\Phi$ - atomic formula.\\

2) $COPY$ operator: $r(COPY(\Phi,n))=r(\Phi)+1$.\\

3) $IF$ operator: $r(IF(\Psi,\Phi_1,\Phi_2))=max\{r(\Phi_1),r(\Phi_2)\}+1$.\\

4) $\alpha$ operator: $r(\alpha(\overline{x}))=max\{r(\Phi_i)\}+1$, where all $\Phi_i\in F$.\\

5) $\Psi$: "$predicate\ P(\overline{x},\overline{y}):\Phi(\overline{x},\overline{y})$"\ , then $r(\Psi)= r(\Phi)+1$.\\ 

6) $\Psi$: "$function\ f(\overline{x})\ return\ y:\Phi(\overline{x},\overline{y})$"\ , then $r(\Psi)= r(\Phi)+1$.\\

7) $P-$D-predicate, then $r(P(\overline{x},\overline{y})) = r(predicate\ P(\overline{x},\overline{y}):\Phi(\overline{x},\overline{y}))+1$\\

8) $f-$D-function, then $r(y:=f(\overline{x})) = r(function\ f(\overline{x})\ return\ y:\Phi(\overline{x},y))+1$\\

9) if $\Phi_1,...,\Phi_n$ - D-formulas, where $max\{r(\Phi_i))\}=k$, then $r(<\Phi_1,...,\Phi_n>)= k+1$.\\ 

\textbf{Lemma 3.1}: If all $\Phi_i-$ boundary $\Delta_0^p-D-$formulas, then 
\begin{center}
$\Phi = <\Phi_1,...\Phi_n>-$ boundary $\Delta_0^p-$formula
\end{center}

$\square$
Induction by $n$:\\
\begin{center}
$D(\mathfrak{M})_E\models <\Phi_1(\overline{x},\overline{y})> \Leftrightarrow D(\mathfrak{M})_{\Gamma_{\Phi_1}(E)}\models true$
\end{center}
and $\Phi_1$ - boundary $\Delta_0^p$-formula.\\

Induction step:\\
Let for any list of length $k$ -  $<\Phi_2,...,\Phi_{k+1}>$ boundary $\Delta_0^p$-formula, then
\begin{center}
$D(\mathfrak{M})_E\models <\Phi_1,...\Phi_{k+1}> \Leftrightarrow D(\mathfrak{M})_{\Gamma_{\Phi_1}(E)}\models <\Phi_2,...\Phi_{k+1}>$
\end{center}

We requare for all $\Phi_i$ - boundary, then for all outcoming $\overline{y}$ for $\Phi_1$ we have $|y_j|\leq C*(|x1|+...|x_n|)^p$, and by induction $<\Phi_2,...,\Phi_{k+1}>$-boundary $\Delta_0^p$ formula, then exists R and q, what $<\Phi_1,...,\Phi_{k+1}>$ is boundary R-q-$\Delta_0^p$ formula. 
$\blacksquare$\\

\textbf{Theorem} Any Delta-program $\Phi$ is $\Delta_0^p$-program.\\ 

$\square$ Proof by induction by formula complexity:\\

Induction Base: $r(\Phi) = 0$: Then our D-formula $\Phi$ is atomic and then it's boundary $\Delta_0^p$-D-formula \\

Induction Step: Let it's true for any $\Phi_i$ with rank $r(\Phi_i)=k$.\\
We have a next:\\

1) $COPY(\Phi,n)=<\Phi,...,\Phi>-$boundary-$\Delta_0^p-$D-formula by definition\\

2) $IF(\Psi,\Phi_1,\Phi_2)-$boundary-$\Delta_0^p-$D-formula.\\

$t(D(\mathfrak{M})_E\models\Psi(\overline{x},\overline{y}))\leq C*(|x_1|+...+|x_n|)^p$ and all $|y_i|\leq C*(|x_1|+...+|x_n|)^p$ and then 
\begin{center}
$t(D(\mathfrak{M})_{\Gamma_{\Psi}(E)}\models\Phi_1(\overline{x}))\leq C*(|y_1|+...+|y_n|)^p\leq n*C^2*(|x_1|+...|x_n|)^{2p}$\\
$t(D(\mathfrak{M})_{E}\models\Phi_2(\overline{x}))\leq C*(|x_1|+...+|x_n|)^p$
\end{center}

3) Operator $\alpha$:
\begin{center}
$t(D(\mathfrak{M})_E\models\alpha(\overline{x}))=t(D(\mathfrak{M})_E\models\Phi_{i(\overline{x})}(\overline{x},\overline{y}))+C*(|x_1|+...|x_n|)^p\leq 2C*(|x_1|+...+|x_n|)^p$
\end{center}

4) $\Phi=<\Phi_1,...,\Phi_n>$ use Lemma 3.1 we get, what $\Phi-\Delta_0^p-$program.\\

5) $\Psi$: "$predicate\ P(\overline{x},\overline{y}): \Phi(\overline{x},\overline{y})$"\\
need check what new symbol $P$ not in signature and $\Phi(\overline{x},\overline{y})$-D-formula\\

6) $\Psi$: "$function\ f(\overline{x}):\ return\ y:\ \Phi(\overline{x},y)$"\\
In this case same argumentation how in 5).\\

7) $P(\overline{x},\overline{y})$, where $P$-D-predicate
\begin{center}
$t(D(\mathfrak{M})_E\models P(\overline{x},\overline{y}))=t(D(\mathfrak{M})_E\models \Phi(\overline{x},\overline{y}))$\\
\end{center}

8) $y:=f(\overline{x})$, where $f$-D-function\\
In this case same argumentation how in 7).\\

$\blacksquare$\\

\textbf{4. Delta methodology in Turing Complete Languages}\\

The process of creating a Delta programs can be transferred to Turing complete languages. In other words, we can talk about the Delta methodology in turing complete languages. All operators, predicates, and functions are was definable for our Delta language are the same as in most programming languages such as PHP, C++, Fortran, Pascal, Solidity. All this programs will be a $p-$computable. Programs using Delta-methodology we can translate on other programming language and also on law language. Delta methodology are important in the direction of  creating smart contracts. For each smart contract, it is very important to know how long it will run and how much computing resources need to be spent for this. Today, the most popular platform for creating smart contracts is Ethereum, and the most popular language is Solidity. And for executing  smart contract on Ethereum need use "gas"\ because we do not know when the contract executing will stop. Delta-methodology for smart contracts decide this questions.\\

In the development of software products it is also very important to know what degree of polynomiality will be our program. Program will be $p-$computable, but the degree of polynomiality is too high. Delta-methodology allows us to estimate the degree of polynomiality in one or another implementation of the algorithm using  programs on high-level languages.

\end{document}